# Grain boundary diffusion and grain boundary structures of a Ni-Cr-Fe- alloy: Evidences for grain boundary phase transformations

Sai Rajeshwari K[a,b], S. Sankaran[a,b], K.C. Hari Kumar[b], Harald Rösner[a], Vladimir A. Esin[c], Sergiy Divinski[a], Gerhard Wilde[a]

[a] Institute of Materials Physics, University of Münster, Germany
[b] Department of Metallurgical and Materials Engineering, IIT Madras, Chennai, India
[c] MINES ParisTech, PSL University, Centre des Matériaux (CMAT), CNRS UMR 7633, Évry, France

Grain boundary tracer diffusion of Ni, Fe and Cr was studied in a Ni-base 602CA coarse-grained alloy. A co-existence of several short-circuit contributions was distinguished at higher temperatures in Harrison's B-type regime (773-873 K), which were related to different families of high-angle grain boundaries with distinct coverages by precipitates and segregation levels as revealed by HAADF-STEM combined with EDX measurements. Annealing at 873 K for 18 hours resulted in $Cr_{23}C_6$-type carbides coexisting with an α-Cr-Mn-enriched phase in addition to sequential segregation layers of Al, Fe and Ni around them. Curved and hackly grain boundaries showed a high density of plate-like carbides, whereas straight grain boundaries were composed of globular carbides with similar chemical composition variations and additionally with alternating layers of Cr and Ni in between the carbides, similar to microstructures after a spinodal decomposition. At lower temperatures, discontinuous interfaces with Cr and Cr-carbide enrichment dominated and the alloy annealed at 403 K for 72 hours contained plate-like $Cr_{23}C_6$-type carbides surrounded by a Ni-rich layer around them. The Ni grain boundary diffusion rates at these relatively low temperatures (formally belonging to C-type kinetics) showed an anomalous character being almost temperature independent. This specific diffusion behaviour was explained by a concomitant relaxation of transformation-induced elastic strains occurring on a longer time scale with respect to grain boundary diffusion. Thermodynamic insights into the probable mechanism of decomposition at grain boundaries are provided.

## 1. Introduction

The alloy 602CA or Nicrofer® 6025 HT is a nickel-base alloy which finds applications in thermal, chemical and petrochemical industries due to its high creep strength and superior corrosion resistance at temperatures as high as 1423 K [1]. In service conditions, this alloy consists of two phases with an austenite matrix and hard carbide precipitates distributed in the matrix as well as at grain boundaries. The presence of a considerable amount of chromium in this alloy leads to the formation of a protective, adherent oxide layer on the exposed surface in a high-temperature oxidizing environment. The formation of slowly growing and non-volatile alumina beneath the chromia scale is proven to enhance the corrosion resistance when compared to other Ni-base alloys, in both cyclic air oxidation and blue flame oil combustion environments [2–4]. Pillai et al. [5] noted that the carbides act as reservoirs which supply Cr for the protective oxide formation at high temperatures (>1273 K) and during long flue gas exposure (>1400 hours). Once the chromium from carbides is consumed, carbon diffuses back to the core (away from the surface), leaving behind a carbide free zone [5]. To comprehend the underlying diffusion kinetics at high-temperatures and in corroding atmospheres, numerous investigations have been undertaken and the element distributions through the depth under the given exposure conditions were established and related to the corrosion process [6–10].

Owing to the nature of the application as a high-temperature corrosion resistance alloy, a comprehensive knowledge of element diffusion is required. $^{51}$Cr and $^{59}$Fe tracer diffusion data in alloys with similar composition are reported in Refs. [11–13] for medium and high temperatures above 673 K. However, the microstructural aspects were not addressed simultaneously, especially with regard to the evolution of the microstructure and the grain boundaries. Moulin et al. [14] evaluated the $^{63}$Ni and $^{51}$Cr diffusion parameters above 1111 K in the 80Ni-20Cr alloy with varied carbon content. A simple comparison of the diffusion data demonstrates that in the temperature range where the carbide precipitates were present, the values of the triple product of Ni diffusion, $P = s \cdot \delta D_{gb}$ (where $s$ is the segregation factor, assumed to be about unity in the alloy, $\delta$ the grain boundary width, and $D_{gb}$ the corresponding grain boundary diffusion coefficient) were enhanced in comparison to the double product $P = \delta D_{gb}$ for Ni diffusion in 99.6 wt.% purity Ni and retarded compared to high-purity 99.999 wt.% Ni [15]. The findings in Ref. [14] were supported by carbide analysis. Unfortunately, there is a lack of diffusion data for moderate and low temperatures where grain boundary diffusion dominates the atomic transport and strongly affects the material properties.

**Table 1**: Chemical composition of Ni-base Alloy 602CA (wt. %)

| Ni | Cr | Fe | Al | C | Ti | Mn | Si | Cu | Zr | Y |
|---|---|---|---|---|---|---|---|---|---|---|
| 62.24 | 25.4 | 9.5 | 2.25 | 0.17 | 0.13 | 0.07 | 0.07 | 0.01 | 0.08 | 0.08 |

In order to fill this gap, the present paper is focused on nickel grain boundary (GB) diffusion in the technologically important 602CA alloy in as wide a temperature interval as possible. Additionally, chromium and iron diffusion in the alloy is studied in the B-type regime after Harrison's classification [16]. Moreover, the transport measurements are here supported by microstructural investigations using analytical transmission electron microscopy (TEM).

## 2. Experimental details
### 2.1 Material

The Ni-base alloy 602CA was procured from ThyssenKrupp® as a solution-annealed and descaled sheet of 6 mm thickness for this study. The nominal chemical composition of the alloy is given in **Table 1**. Disk specimens with 3 mm thickness and 8 mm in diameter were cut using spark erosion. One face of each disc was polished to a mirror-like finish using standard metallographic techniques and the final polishing was performed using an OP-S standard colloidal silica suspension from Struers®. The discs were subjected to a solution treatment in a vacuum furnace at 1173 K for 16 hours in order to eliminate the preparation-induced defects and also to stabilize the microstructure before tracer application.

### 2.2 Microstructure characterization

The specimens were prepared using standard metallographic techniques. Additionally, for an electron back-scatter diffraction (EBSD) analysis samples were subjected to OP-S colloidal silica polishing using a Tegramin Struers® automatic polisher. The microstructural examination was performed by a FEI Nova NanoSEM 230 scanning electron microscope (SEM) operating at 15 to 20 kV. EBSD-based orientation imaging microscopy (OIM) was used to obtain the local crystallographic orientation information and the data were analysed using the TSL software. The step size used for the EBSD measurements was 0.5 µm.

For TEM analyses, specimen approximately 100 µm in thickness were first prepared by mechanical thinning (grinding). Then, electron transparent regions were obtained by electro-polishing at room temperature using a mixture of perchloric- and acetic acids at 1:9 ratio at 14 V using Struers® Tenupol-5. TEM preexamination was carried out with a Zeiss Libra 200FE operating at an accelerating voltage of 200 kV. Analytical TEM measurements were performed with a Titan Themis G3 60-300 TEM operated at 300 kV and equipped with FEI's ChemiSTEM technology [17].

### 2.3 Radiotracer diffusion experiments

The polished discs were pre-annealed first at the conditions of the subsequent diffusion annealing treatments to ensure almost equilibrium conditions for the intended grain boundary diffusion measurements, at least for the B-type kinetic conditions.

The $^{63}$Ni radioisotope (β decays, half-life of 100 years) was available as a 0.5M HCl solution. It was diluted with double-distilled water to obtain a specific radioactivity of about 1 kBq µl$^{-1}$. Few microliters of the tracer solution with a total activity of 6-8 kBq were dropped on the polished disc surface and allowed to dry. These samples were sealed in evacuated quartz ampoules, filled with purified Ar and annealed at selected temperatures. The same experimental procedure was applied for the experiments on Fe and Cr diffusion using the $^{59}$Fe and $^{51}$Cr isotopes (γ-decays, half-lives of 45.1 and 27.7 days, respectively).

After the diffusion annealing treatments, the diameter of the samples was reduced by at least 1 mm to eliminate probable contributions of lateral and surface diffusion. Parallel sectioning was performed using precision mechanical grinding and simultaneous weight measurements for the layer thickness evaluation. The penetration depth was determined from the layer thicknesses obtained by measuring the disc mass before and after sectioning with a relative accuracy better than 0.3%.

The radioactivity of the γ-decaying elements was precisely measured by a CANBERRA well-type intrinsic Ge γ-detector equipped with a 16 K multi-channel analyser discriminating the corresponding emission energies. The concentration profiles for the β-decaying $^{63}$Ni isotope were measured by a Liquid Scintillation Counter (LSC) TRI CARB 2910 TR. The measured background-subtracted counting rates divided by the section mass are proportional to the section-averaged tracer concentration [18].

## 3. Results
### 3.1. Microstructure investigation

Figure 1 shows the microstructure (with color-coded GBs and a superimposed EBSD image quality map) of the solution-annealed material at 1173 K for 16 hours. The average grain size is determined to be 45 ± 9.5 μm (counting the twin boundaries as grain boundaries).

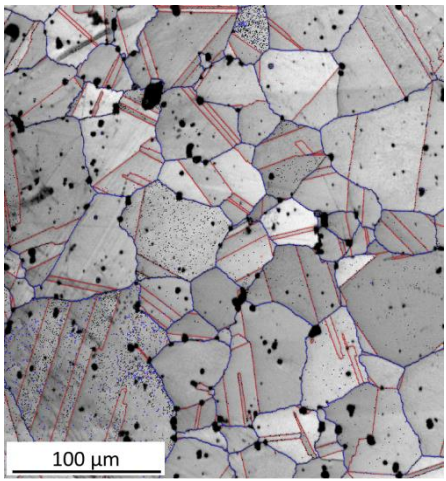

**Figure 1**: General high-angle grain boundaries (blue), low-angle grain boundaries (yellow) and Σ3 coherent twin boundaries (red) superimposed on the image quality map obtained from EBSD for 602CA alloy after solution annealing at 1173 K for 16 hours. The unindexed black regions correspond to lumpy carbide precipitates.

As one can see, the majority of boundaries were high-angle grain boundaries (HAGBs), among which numerous recrystallization-induced Σ3 twin boundaries were observed. The fraction of low-angle grain boundaries (LAGBs) with the misorientation angle < 15°, of Σ3 twin boundaries and HAGBs were approximately 0.13, 0.33 and 0.54, respectively. By inspecting the microstructure, Fig. 1, we noted that no single segment of a LAGB is longer than 20 μm, thus general (random) HAGBs (blue lines) should provide the dominant contribution to short-circuit diffusion paths for the atomic transport in the material in the present conditions for penetration depths larger than 20 μm.

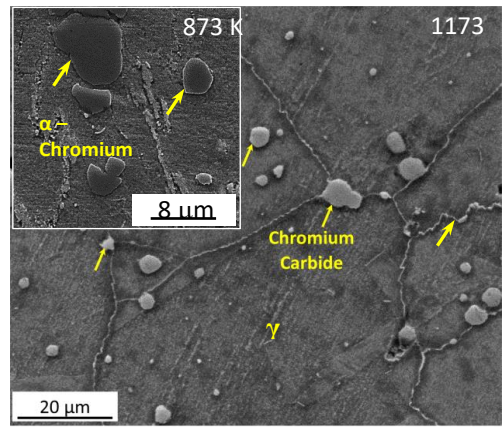

**Figure 2**: Scanning electron micrographs of alloy 602CA after solution annealing at 1173 K for 16 hours, showing the morphology and abundance of chromium carbides in the grain interiors, at grain boundaries, and at triple junctions. The inset shows the presence of additional chromium-rich bcc phase (α-Cr) after annealing at 873 K for 18 hours

The presence of high amounts of chromium and carbon in the alloys led to the formation of mainly $M_{23}C_6$ carbides (the unindexed regions in the image quality map shown in Figure 1). The globular morphology of carbides and their distribution at grain boundaries and triple junctions are shown in Figure 2. In the present solution annealed condition, the carbides are of the $Cr_{23}C_6$ types as it follows from the previous thermodynamic calculations using the JMATPRO® 6.2.1 software and electron probe micro-analysis (EPMA) measurements [5]. Our equilibrium calculations using the Thermo-Calc software and the TCNI8 thermodynamic database [19] shows that at 1173 K, the equilibrium phases are γ (96.2 vol %), (Cr,Fe,Ni)$_{23}$C$_6$ (3.3 vol %) and Ni$_{17}$Y$_2$ (0.5 vol %). The γ' solvus is at 1061 K and Cr-rich bcc begins to form at 1009 K. At 873 K the microstructure consists of γ (74 vol %), γ' (15.8 vol %), Cr-rich bcc (6.3 vol %), (Cr,Fe,Ni)$_{23}$C$_6$ (3.4 vol %) and Ni$_{17}$Y$_2$ (0.5 vol %). It is also seen from the calculations that there is very little solubility of Mn in $M_{23}C_6$. Moreover, the area fraction of the granular carbides in the alloy is around 3% as estimated from SEM images and correlates with the equilibrium volume fraction of $M_{23}C_6$ obtained using JMATPRO® 6.2.1 [5] and Thermo-Calc. TEM investigation on the alloy annealed at these two temperatures confirms the ubiquitous presence of $M_{23}C_6$ at 1173 K and 873 K and documents the appearance of a chromium-rich bcc phase, α-Cr, evolving in significant fractions after annealing at 873 K (Fig. 3).

The globular carbide precipitates were larger at grain boundaries, typically several μm in diameter,

some even reaching 10 μm in size, while smaller precipitates were seen in the crystalline bulk (Fig. 2). The majority of general HAGBs were found to be decorated by carbides. The major metal component in these carbides at room temperature was chromium. However, some fractions of nickel, manganese, aluminium and iron were recorded, too, either dissolved in or segregated at the precipitates, depending on the annealing temperature.

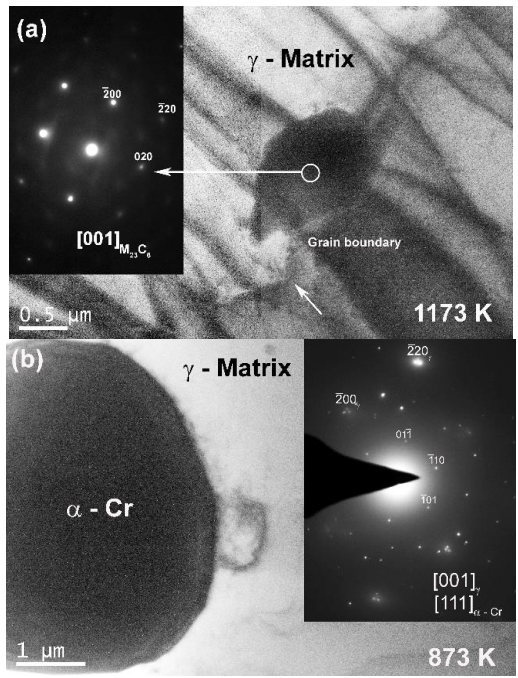

**Figure 3:** Bright field TEM micrographs and the corresponding micro diffraction patterns (insets) exhibiting (a) coarse $Cr_{23}C_6$ carbide precipitate in a solution annealed sample at 1173 K for 16 hours (b) a chromium rich bcc phase, α – Cr, which evolves after annealing at 873 K for 18 hours.

Figures 4a and 4b represent high-angle annular dark-field (HAADF) images showing triple junction areas decorated with precipitates and also possibly element segregation after annealing treatments at 403 K and 873 K for 72 and 18 hours, respectively. The precipitation layer (mainly as $M_{23}C_6$ carbides) is, however, not continuously distributed over the grain boundaries after low-temperature annealing (Fig. 4a) and strongly segmented GBs with a high density of structural features that resemble kinks or disconnections were seen at elevated temperatures (Fig. 4b). A detailed elemental analysis of these precipitates and the concurrently present segregation fields are presented in the following section in conjunction with the diffusion phenomena along grain boundaries at two different annealing temperatures. In contrast to the conditions present at elevated temperatures, a high density of dislocation segments residing directly at grain boundaries is visible at lower temperatures, indicating local strains, c.f. Figs. 4a and 4b. These features of the grain boundary structure turned out to be critical for the analysis of the measured diffusion profiles since dislocations present further short-circuit diffusion paths and their contribution has to be carefully evaluated (see below).

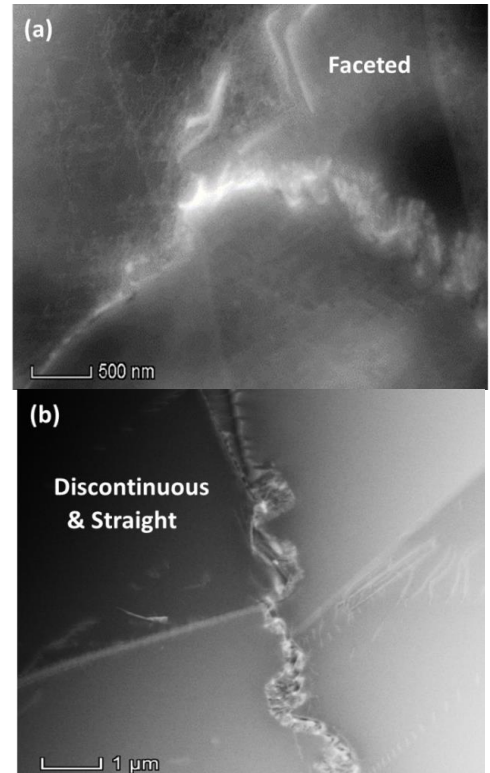

**Figure 4**: HAADF-STEM images showing GB triple junctions after annealing (a) at 403 K for 72 hours and (b) at 873 K for 18 hours. Grain boundaries decorated with precipitates are visible.

### 3.2. Radiotracer measurements

The diffusion mechanisms and the developing penetration profiles for the diffusing atoms depend on a number of critical parameters such as annealing temperature, grain boundary segregation, diffusion time and microstructure. Harisson's classification scheme [16] has been elaborated for grain boundary diffusion in polycrystals, providing the required basis for a proper analysis of the measured concentration profiles in different kinetic regimes. The B- and C-type regimes are relevant for our study and the values of the corresponding Le Claire parameters α and β [20],

$$\alpha = \frac{s\delta}{2\sqrt{D_v t}} \qquad (1)$$

$$\beta = \frac{P}{2D_v\sqrt{D_v t}} \qquad (2)$$

are of prime importance. Here $D_v$ is the coefficient of bulk (volume) diffusion, $P = s\delta D_{gb}$ is the triple product (see Introduction), and $t$ is the diffusion time. The diffusional grain boundary width $\delta$ was measured for a number of FCC materials to be 0.5 nm [21–24] (for an in-depth discussion see e.g. Ref. [25]). Note that the parameters $\alpha$ and $\beta$ are given here for the case of solute diffusion when the segregation factor $s$ is taken into account [25]. Since Ni, Cr and Fe are the main alloying elements in the material, we are not expecting their strong segregation to the grain boundaries. A moderate enrichment (or depletion) of these elements at the grain boundaries, which is in fact found in the present material, would correspond to the value of $s$ less than about two or three. Thus, we can safely assume that $s \approx 1$ for all tracer elements in the order-of-magnitude estimates using Eqs. (1) and (2).

Furthermore, in all diffusion experiments the bulk diffusion length, $2\sqrt{D_v t}$, was chosen to be significantly smaller than the grain size, thus, strictly keeping the requirements for the B-type kinetic regime even at elevated temperatures [16,25].

One has to highlight that the solute segregation factor in the GB diffusion problem is defined as the ratio of the solute concentration at a GB, $c_{gb}$, to that in the adjacent crystalline bulk, $c_v^0$, $s = c_{gb}/c_v^0$ [24]. In the present case, both tracer and solute atoms of the same chemical element contribute to the thermodynamic equilibrium at the interface [26]. Bernardini and co-workers [27,28] have shown that the equilibrium segregation is established shortly after commencement of a B-type diffusion measurement and the total concentration of the given chemical element has to be taken into account to determine the value of $s$. Moreover, since the tracer concentration is typically very small in a radio-tracer experiment, it can be neglected and it is the solute concentration in the alloy which determines the equilibrium segregation factor [29].

The C-type kinetic conditions are usually satisfied for low temperatures and short diffusion times, where diffusion in the bulk is negligible and diffusional transport solely proceeds along grain boundaries. These conditions correspond to $\alpha > 1$ in Eq. (1) and the grain boundary diffusion coefficient, $D_{gb}$, can directly be determined from the concentration profile. If the Gaussian solution can be applied, $D_{gb}$ is proportional to the slope in the $ln\bar{c}$ vs $y^2$ plot,

$$D_{gb} = -\frac{1}{4t}\left(\frac{\partial ln\bar{c}}{\partial y^2}\right)^{-1} \qquad (3)$$

Here $\bar{c}$ and $y$ are the tracer concentration and the penetration depth, respectively.

The B-type regime prevails typically at higher temperatures when the volume diffusion length is significantly larger than the grain boundary width. However, the diffusion fluxes from different grain boundaries should simultaneously not overlap, i.e., $\alpha \ll \sqrt{D_v t} < \frac{d}{3}$, $d$ is the grain size [25]. In such conditions, the grain boundaries can be treated as isolated entities and Le Claire's [20] solution for instantaneous (Suzuoka) [30] or constant (Whipple) [31] sources can be applied. The B-type kinetics, thus, corresponds to the $\alpha < 0.1$ and $\beta > 10$ conditions and the triple product $P$ is obtained from the linear fit of the grain boundary diffusion-related branches of the penetration profiles in the coordinates of $ln\bar{c}$ vs $y^{6/5}$,

$$P = 1.308\sqrt{\frac{D_v}{t}}\left(-\frac{\partial ln\bar{c}}{\partial y^{\frac{6}{5}}}\right)^{-\frac{5}{3}} \qquad (4)$$

In the present estimates we used the values of the volume diffusion coefficients, $D_v$, measured by Maier et al. [32] for Ni

$$D_v = 9.2 \times 10^{-5} \exp\left(-278 kJ \frac{mol^{-1}}{RT}\right) \text{ m}^2\text{s}^{-1} \qquad (5)$$

and Million et al. [33] for Fe

$$D_v = 3.8 \times 10^{-4} \exp\left(-298 kJ \frac{mol^{-1}}{RT}\right) \text{ m}^2\text{s}^{-1} \qquad (6)$$

and Cr

$$D_v = 2 \times 10^{-4} \exp\left(-284.5 kJ \frac{mol^{-1}}{RT}\right) \text{ m}^2\text{s}^{-1} \qquad (7)$$

Here $R$ is the universal gas constant and $T$ the diffusion temperature.

### 3.2.1. C-type kinetic regime of Ni grain boundary diffusion

The C-type Ni diffusion measurements in the coarse-grained 602CA alloy were performed at three different temperatures: 403, 573 and 673 K. As expected, larger diffusion depths were observed at higher diffusion annealing temperatures in the C-type regime (Fig. 5a). As it was discussed above, the C-type measurements directly provide the grain boundary diffusion coefficient, $D_{gb}$, see Eq. (3).

The obtained experimental profiles suggest the existence of two short-circuit diffusion paths. However, the near-surface (relatively slow) diffusion paths (penetration depths < 2 μm) could potentially be affected by the sample preparation and mechanical grinding procedure; thus, we did not analyse them for the C-type kinetic data. The experimental parameters and the determined diffusion coefficients of the main (fast) diffusion paths are listed in Table 3. The determined diffusion coefficients are plotted below in Fig. 8a in the Arrhenius coordinates.

### 3.2.2. B-type kinetic regime of Ni grain boundary diffusion

In the B-type kinetic regime, GB diffusion measurements were performed in the as-received, solution-annealed alloy 602CA (initially annealed at 1173 K for 16 h) at three temperatures and the corresponding penetration profiles are shown in Fig. 5b.

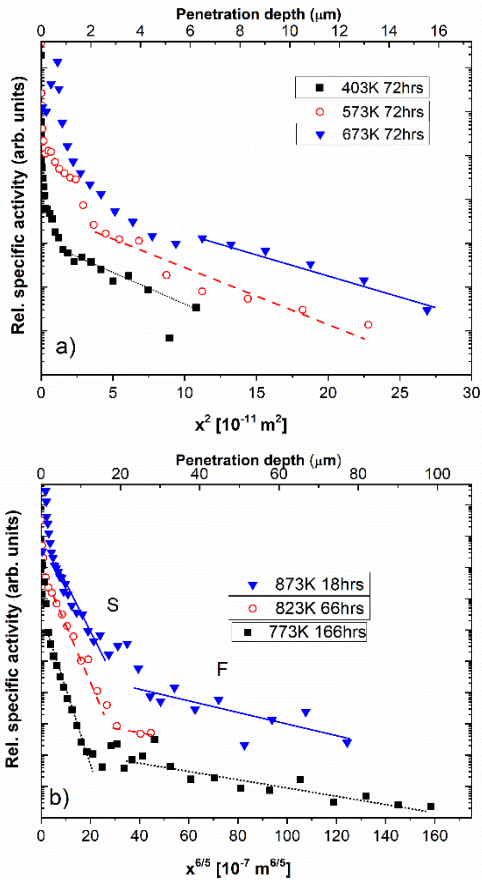

**Figure 5**: Ni tracer diffusion penetration profiles measured in the C-type (a) and B-type (b) regime in the as-received solution annealed alloy (pre-annealed at 1173 K for 16 h). Near-surface (< 2 μm) branches are affected by the grinding procedure and by direct volume diffusion at elevated temperatures and they are neglected in the analysis. Slow (S) and fast (F) short-circuit diffusion paths are distinguished.

The near-surface points at depths below 2 μm correspond to direct diffusion into the crystalline bulk (the bulk diffusion length, $2\sqrt{D_v t}$, is a fraction of a micrometre for all these experiments, see Tables 3 and 4). However, they are not included in the subsequent analysis since the number of relevant data points is too small to provide any reliable estimates and, as it was mentioned above, this branch could potentially be affected by the mechanical grinding procedure.

Short-circuit diffusion contributes to the tracer penetration to larger depths and two such branches can clearly be distinguished, with steeper and shallower slopes in the pertinent coordinates of Fig. 5b. Since the corresponding diffusion rates are inversely proportional to the slopes of the linear fits in the given coordinates, see e.g. Eq. (4), one may call the two branches slow (S) and fast (F) branches. These systematics assume, e.g., that two basically different populations of short circuits exist in the material which act simultaneously, providing significantly different rates of atomic transport along them. Similar features were previously observed for, e.g., Zn diffusion in hydro-extruded Al which were attributed to the coexistence of grain and sub-grain (low-angle) boundaries in the material [34]. This is definitely not a unique explanation for the existence of slow and fast diffusion paths on top of the common diffusion path via the crystalline bulk. One may think in terms of either (I) dislocation and grain boundary diffusion contributions, though their mutual arrangement and resulting concentration profiles can be very sophisticated [35] or (II) an arrangement of two different types of grain boundaries as distinct short circuit paths. The latter situation has been observed and carefully analysed for different materials from sintered nanocrystalline [36, 37] to ultrafine grained ones processed by severe plastic deformation [38-42]. Below we will analyse both these hypotheses.

**I. Pipe diffusion as a slow-diffusion branch**. Within this interpretation, we are assuming that a network of dislocation pipes represents the slow diffusion paths and general high-angle grain boundaries provide for fast diffusion. According to the Le Claire-Rabinovich solution [43], diffusion along dislocation pipes results in concentration profiles with the logarithm of the layer tracer concentration decaying linearly with the penetration depth. The corresponding triple product, $P_d = s_d a^2 D_d$, was determined from the pertinent slope (which is time-independent). Here we are extending the treatment of Le Claire and Rabinovich including the solute segregation at dislocations which is characterized by the segregation factor $s_d$. $D_d$ is the dislocation diffusion coefficient and $a$ is the radius of the dislocation pipes (in the numerical estimates we will use $a \approx \delta$ and the segregation factor at dislocations, $s_d$, is assumed to be about unity).

Considering the two contributions, i.e. dislocation and grain boundary diffusion branches, the experimental profiles have to be fitted by a sum of two exponential functions, Eq. (8a), that allows determining both, the dislocation, Eq. (8b), and grain boundary triple products, Eq. (4).

$$c(y) = P_d + P_F = A_1 \exp(-g_1 * y) + A_2 \exp\left(-g_2 * y^{\frac{6}{5}}\right) \quad (8a)$$

$$P_d = s_d a^2 D_d = A^2 D_v g_1^{-2} \quad (8b)$$

Here $A_1$, $A_2$, $g_1$, and $g_2$ are fitting parameters; $A$ is a numerical factor which depends on the value of $\alpha$ and varies typically between 0.5 and 2 [43]. In the present conditions, it was estimated to be about 0.6. The values evaluated for $P_d$ and calculated for $D_d$ from Eq. (8b) that are associated with dislocation pipe diffusion for Ni, Fe and Cr tracers are listed in Table 2.

**Table 2:** Triple product of dislocation diffusion, $P_d$, and the evaluated pipe diffusion coefficient, $D_d$, for Ni, Fe and Cr tracer diffusion, from the calculations of B-regime penetration profiles according to Eqs. (8a) and (8b).

| T (K) | t ($10^4$ s) | Ni $P_d$ ($10^{-34}$ m$^4$/s) | Ni $D_d$ ($10^{-16}$ m$^2$/s) | Fe $P_d$ ($10^{-34}$ m$^4$/s) | Fe $D_d$ ($10^{-16}$ m$^2$/s) | Cr $P_d$ ($10^{-34}$ m$^4$/s) | Cr $D_d$ ($10^{-16}$ m$^2$/s) |
|---|---|---|---|---|---|---|---|
| 873 | 6.48 | 59.4 | 237.6 | 6.8 | 27.2 | 29.43 | 117.72 |
| 823 | 23.76 | 3.54 | 14.16 | 0.28 | 1.12 | 0.92 | 3.68 |
| 773 | 59.76 | 0.11 | 0.44 | | | | |

We found that the tentatively determined diffusion coefficients $D_d$ reveal an Arrhenius temperature dependence. However, the values become unreasonably low with respect to the GB diffusion coefficients, especially at 773 K. The estimated ratio $D_d/D_{gb}$ was less than $10^{-5}$ at 773 K, although values of about 0.01 - 0.1 are generally to be expected [25] as it was, e.g., determined for Ag GB self-diffusion [22]. This fact provides strong hints against a dominant contribution of the dislocation pipes as a separate short-circuit path corresponding to the slow diffusion branches in Fig. 5b.

The TEM data, see e.g. Fig. 6, suggest the existence of dislocation bands with a high dislocation density (of approximately $10^{13}$-$10^{14}$ m$^{-2}$), separated by distances larger than several micrometres. The diffusion data suggest that the dislocation density was probably not high enough to provide a distinct contribution to the penetration profiles, which would be measurable in the present conditions.

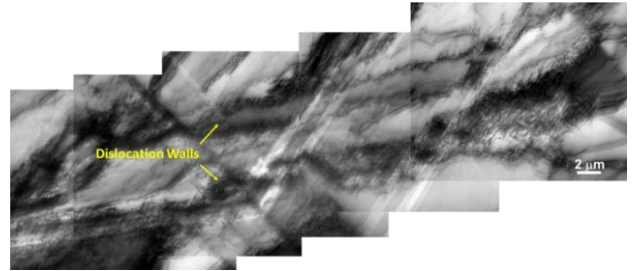

**Figure 6**: Bright-field TEM panoramic image displaying long continuous dislocation walls and networks sheared by the large twins present in the solution-annealed condition of the Ni-base alloy.

Hence we conclude that two grain boundary types exist, which cause the appearance of two branches in the concentration profiles. The following microstructure analysis will support this interpretation, i.e. the existence of two distinct types of GBs acting as slow and fast diffusion paths.

**II. Two types of GBs as slow and fast diffusion branches.** The penetration profiles are fitted as

$$c(y) = P_d + P_F = A_1 \exp\left(-g_1 \times y^{\frac{6}{5}}\right) + A_2 \exp\left(-g_2 \times y^{\frac{6}{5}}\right) \quad (9)$$

and the corresponding triple products are determined by the pertinent slopes, i.e. by the values of $g_1$ and $g_2$ using Eq. (4). The determined triple products are listed in Table 3.

By using Equation (9) we assume that the two branches provide two parallel and independent contributions to the total atomic transport. The analysis below, see Discussion part, supports this assumption.

### 3.3. B-type regime of Fe and Cr GB diffusion

The penetration profiles measured for Fe and Cr diffusion in the B-type regime are shown in Fig. 7. Since both Fe and Cr are alloying elements in the material, the grain boundary diffusion can be measured in a single combined experiment. The very small amount of the tracers applied to the surface does not affect the chemical composition and does not modify the chemical interaction at grain boundaries. Note that this would not be the case for impurity diffusion of Fe and Cr in, e.g., pure

**Table 3**: Diffusion parameters (temperature $T$, time $t$, and the values of the Le Clair parameters α and β) and the evaluated triple products ($P_F$ & $P_S$) and GB diffusivities ($D_{gb}$) for Ni diffusion. The values of $P_F$ and $P_S$ correspond to the fast and slow diffusion paths, respectively.

| $T$(K) | $t$ ($10^4$ s) | $D_v$ (m²/s) | $\sqrt{(D_v t)}$ (nm) | $P_{S-Ni}$ ($10^{-26}$ m³/s) | $\beta_{S-Ni}$ ($10^4$) | $P_{F-Ni}$ ($10^{-24}$ m³/s) | $\beta_{F-Ni}$ ($10^6$) | $D_{gb}$ ($10^{-17}$ m²/s) | α | Regime |
|---|---|---|---|---|---|---|---|---|---|---|
| 873 | 6.48 | 2.14e-21 | 11.77 | $458.6^{+53.1}_{-53.1}$ | 8.89 | $68.3^{+4.6}_{-21.5}$ | 6.24 | - | $2.1 \times 10^{-2}$ | B |
| 823 | 23.76 | 2.09e-22 | 7.04 | $44.84^{+3.98}_{-3.98}$ | 14.87 | $17.8^{+8.36}_{-7.12}$ | 5.9 | - | $3.5 \times 10^{-2}$ | B |
| 773 | 59.76 | 1.51e-23 | 3.00 | $7.69^{+1.08}_{-1.08}$ | 82.67 | $6.76^{+0.83}_{-3.21}$ | 72.7 | - | $8.3 \times 10^{-2}$ | B |
| 673 | 25.92 | 2.43e-26 | 0.08 | - | - | - | - | $4.89^{+1.5}_{-0.82}$ | 3.2 | C |
| 573 | 25.92 | 4.18e-30 | 0.001 | - | - | - | - | $2.7^{+0.69}_{-0.55}$ | $2.4 \times 10^2$ | C |
| 403 | 25.92 | 8.53e-41 | ~$10^{-8}$ | - | - | - | - | $2.13^{+0.06}_{-0.03}$ | $5.3 \times 10^7$ | C |

Ni which in fact would correspond to co-diffusion of the elements.

The penetration depth profiles obtained for Fe and Cr tracer diffusion were much shorter and similar in their trends to the diffusion profiles for the Ni tracer. Contributions of both, slow- and fast-diffusion paths can be distinctly recognized in Fig. 7, similar to the case of Ni diffusion. The triple products, $P_{S-Fe}$, $P_{F-Fe}$ and $P_{S-Cr}$, $P_{F-Cr}$ and the diffusion parameters α and β are listed in Table 4.

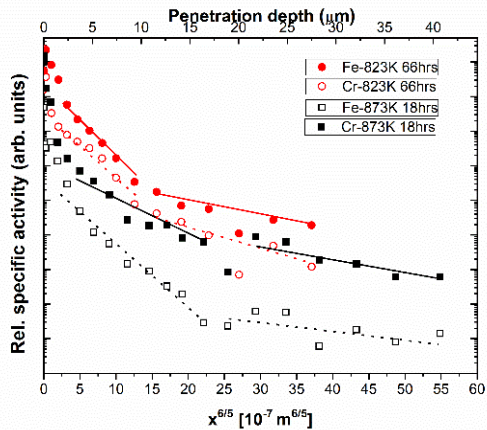

**Figure 7**: Fe and Cr tracer diffusion penetration profiles in the B-type regime of as-received solution-annealed alloys. Both the profile branches (referring to slow and fast GB diffusion paths) are linear in the given coordinates.

### 3.4. Temperature dependence of GB diffusion

The determined diffusivities, i.e. the triple products and the diffusion coefficients for the B and C-type measurements, respectively, for Ni GB diffusion are plotted as a function of the inverse temperature in Fig. 8a. Analysis proves that in the case of Ni diffusion at all three temperatures proceeds in the intended B-type regime α < 0.1 and β >> 10, thus diffusion corresponds to the true B-type case, see Table 3.

**Table 4**: Diffusion parameters (α, β) and the evaluated triple products ($P_{S-Fe}$, $P_{F-Fe}$ and $P_{S-Cr}$, $P_{F-Cr}$) for Fe and Cr diffusion in the B-type regime.

| $T$ (K) | $t$ ($10^4$ s) | $D_v$ ($10^{-23}$ m²/s) | $P_S$ ($10^{-26}$ m³/s) | $\beta_S$ ($10^5$) | $P_F$ ($10^{-24}$ m³/s) | $\beta_F$ ($10^6$) | α |
|---|---|---|---|---|---|---|---|
| Fe tracer | | | | | | | |
| 873 | 6.48 | 57.23 | $80.57^{+11.1}_{-11.1}$ | 1.754 | $29.2^{+29.2}_{-12.8}$ | 0.09 | 0.04 |
| 823 | 23.76 | 4.73 | $7.08^{+0.88}_{-0.88}$ | 3.782 | $1.73^{+0.81}_{-0.41}$ | 7.05 | 0.08 |
| Cr tracer | | | | | | | |
| 873 | 6.48 | 18.98 | $33.7^{+6.93}_{-6.93}$ | 1.399 | $15.07^{+1.75}_{-4.33}$ | 0.13 | 0.02 |
| 823 | 23.76 | 1.75 | $1.29^{+0.16}_{-1.290.16}$ | 1.135 | $3.16^{+1.25}_{-0.88}$ | 3.85 | 0.04 |

## 4. Discussion
### 4.1. Slow and fast short-circuit diffusion paths

Grain boundary diffusion of Ni in the as-received 602CA alloy was measured in a wide temperature interval under both B and C-type kinetic conditions, while Fe and Cr diffusion was measured only in the B-type kinetic regime. Two short-circuit diffusion contributions could systematically be recognized for all measured B-regime profiles, a fact which suggests the existence of two families of short-circuit paths. The detailed analysis of diffusion data, Tables 2 and 3, suggests that these are two families of grain boundaries which provide the two distinct contributions.

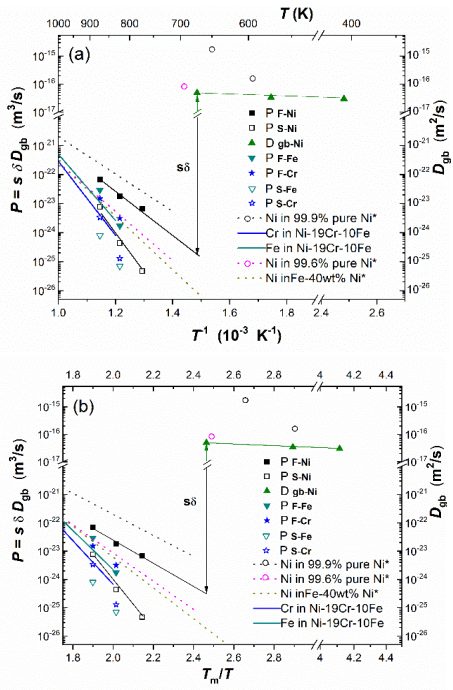

**Figure 8**: a) The triple products $P$ (left ordinate) and the diffusion coefficients (right ordinate) measured in the B- and C-type regimes, respectively, as functions of the inverse temperature. The letters F and S distinguish between the diffusivities along the fast & slow diffusion paths. Ni grain boundary diffusivities in polycrystalline nickel of different purity levels [15], are shown by dotted lines in the B-type regime and open circles in the C-type regime. The Ni triple product in nano-GBs of Fe-40wt%Ni is depicted as dotted orange coloured line [44]. The measured triple products for Fe (inverted green triangles) and Cr (blue stars) are shown. b) The triple products measured in the B-type regimes of the corresponding materials as functions of the inverse homologous temperature.

It should be noted here that there is a broad spectrum of grain boundaries with different diffusion properties which were most probably randomly distributed in the investigated samples. However, the shapes of the penetration profiles favour adominance of two families of such interfaces with a bimodal distribution of the pertinent diffusion properties. In the analysis which follows we will use the approximation corresponding to the replacement of the potentially continuous distribution by two representative "effective" interface types. The entire amount of the data confirms the validity of such an approximation.

Potentially, at least two different arrangements of the short-circuit paths may be suggested, which implies two different theoretical frameworks for the data analysis:

I) A hierarchical arrangement of the short-circuit paths
II) Two independent short-circuit paths which act in parallel (assuming thus no diffusional interaction of the two paths, i.e. no exchange of tracer atoms between these paths);

In case (I) there is a continuous exchange of tracer atoms between the short-circuit paths and, e.g., one tracer atom inspects several fast paths using slow paths as short-circuits between them. The arrangement (I) was found, e.g., in the case of nanocrystalline Fe-Ni alloys [44], severely plastically deformed copper [38] or polycrystalline α-Fe [35]. The analysis of the concentration profiles becomes very involved, and the Harrison classification was extended to include the hierarchy of the short-circuit paths [36,37,44]. According to the classification suggested by Divinski [37] the conditions of so-called C-C and C-B regimes might be satisfied in the present case, for a detailed description see [25]. The key parameter of the arrangement (I) is the relative density of the slow short-circuit paths crossing the fast diffusion paths and providing a leakage from the latter in a typical diffusion experiment [44]. We found that in the present case both the density of low-angle GBs or dislocations, which might act as slow-diffusion paths is too low to provide any significant contribution for the case of *hierarchically arranged paths*. For example, a dislocation density of about $10^{15}$ m$^{-2}$ would be required to induce the diffusion conditions observed here, which is definitely not the case in the solution-annealed alloy. Thus, dislocations did not affect the kinetics of GB diffusion.

In case (II), there exists no exchange of the tracer atoms between the two independent paths. This means that the distance between the paths was larger than the diffusion length in the bulk. In the present conditions, this means that the two paths have to be separated by at least 3 μm. An inspection of the microstructure shown in Fig. 1 suggests that the simultaneous contributions of different types of GBs forming independent networks that act as two parallel (i.e. there is no tracer exchange between the short circuits) short circuit diffusion pathways would satisfy this criterion.

The existence of the two branches for short-circuit diffusion was unambiguously revealed at higher temperatures for the B-type measurements. Their presence cannot be ruled out at lower temperatures, too, see Fig. 5a, however a reliable determination of the pertinent diffusion characteristics for the slow branch was not possible in view of the limitations of the used mechanical sectioning technique with respect to the depth resolution.

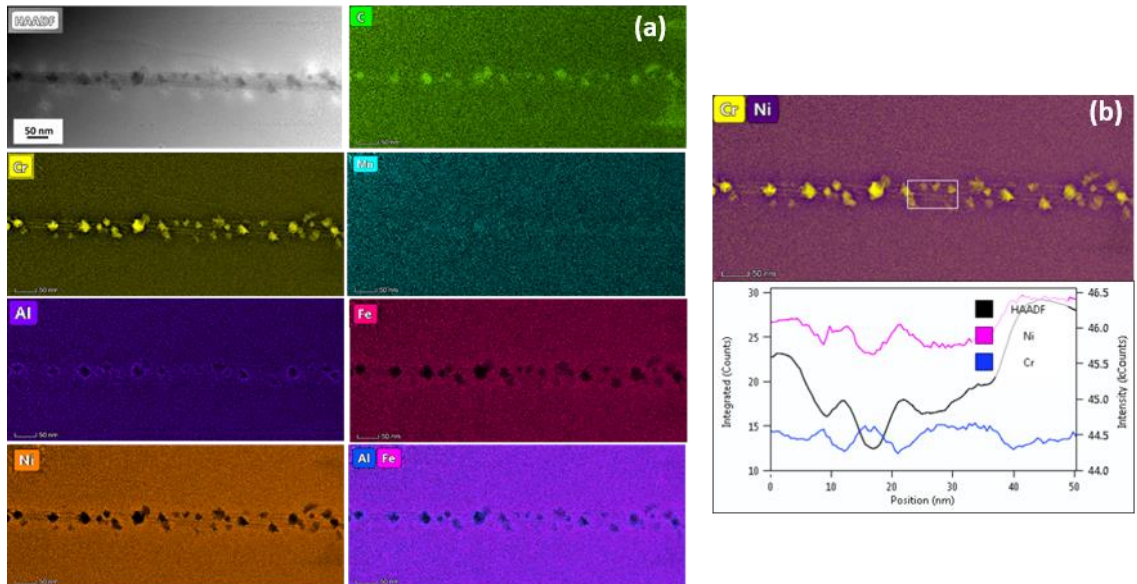

**Figure 9:** (a) HAADF-STEM image of alloy 602 CA annealed at 873 K for 18 hours showing a straight GB and the corresponding elemental EDS maps. The high-angle boundary was inclined to the beam direction and contains numerous particles. The grain boundary globular particles are rich in Cr, C and small amounts of Mn. The multiple layers around the particles display successive elemental segregation of (counting outwards from a particle) Ni, Fe and Al, see the Al-Fe map together. (b) Overlapped Cr and Ni elemental EDS maps for the area shown in (a). Also plotted is the intensity along the line (inside indicated white box) drawn perpendicular to the boundary containing the globular carbide and Cr-Ni enriched/depleted straight lines. Note that these straight lines appear over a width of around 50 nm. It can be seen that the HAADF intensity coincides with the Ni concentration profile. However, the Cr concentration maxima coincide with the Ni concentration minima, suggesting alternating Ni and Cr layers.

Below we will present the result of TEM inspection of the GB structures and chemistry with a proof of the existence of different families of high-angle grain boundaries.

First of all, we have to underline that these are two basically different – but complementary – techniques, namely tracer diffusion measurements and electron microscopy. While tracer diffusion provides integral data for the whole sample, the TEM analysis is highly local. However, we do observe two fundamentally different types of high-angle grain boundaries with respect to GB structure, segregation and precipitation that in view of the present kinetic data allows important generalizations.

Secondly, the existence of the two families of general high-angle GBs with two different diffusivities is definitely a crude approximation and there exist a whole spectrum of interfaces with varying diffusivities. However, the linearity of the two branches of the penetration profiles in the corresponding coordinates could be followed over three to four and over two to three orders of magnitude for slow and fast paths, respectively, with a relatively sharp transition between them (see Fig. 5). This experimental fact supports the relevance of the two-branch approximation.

### 4.2 Microstructural evidence for grain boundary segregation and phase transformations

The variations of the chemical composition along the grain boundaries and in the matrix were investigated using HAADF-STEM combined with a four-quadrant EDS detector system. In order to determine the characteristics of the GBs related to the diffusion path, the grain boundaries were examined in detail after an annealing treatment at 873 K for 18 hours as well as after annealing at 403 K for 72 hours, i.e. associated with the B- (higher temperatures above about 700 K) and C-type (lower temperatures below about 700 K) regime, respectively.

#### 4.2.1 GB structures observed at higher temperatures (>700 K)

TEM inspection revealed the existence of two dominant types of high-angle GBs; that is (i) straight ones with globular carbide precipitates and (ii) curved, hackly ones with plate-like precipitates.

Figure 9a shows HAADF-STEM images and the corresponding EDS maps of straight grain boundaries in the Ni-base alloy after annealing at 873 K for 18 hours. In addition, combined Cr and Ni elemental maps are shown in Fig. 9b that correspond to the regions displayed in the HAADF images. An example of curved, highly broken GBs is shown in Fig. 10.

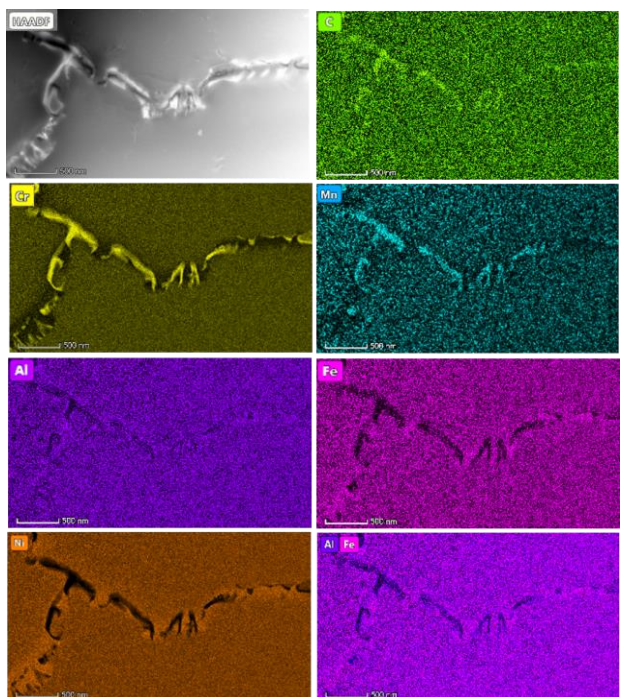

**Figure 10:** HAADF-STEM image of alloy 602 CA annealed at 873 K for 18 hours showing a hackly GB and the corresponding elemental EDS maps. The plate-like particles at the grain boundary were rich in Cr, C and Mn. Note that the multiple layers around the particles display successive elemental segregation of Al, Fe and Ni. The first layer around the particles was Al-rich, followed by Fe- and finally Ni-rich layers (see the Al-Fe combined elemental map).

The globular carbide precipitates (dominantly Cr carbides) at the straight boundaries are significantly smaller and morphologically different from the plate-like Cr carbides detected at the curved grain boundaries. Two distinct features should be mentioned. Firstly, these particles are not only rich in Cr and C but rich in Mn, too, especially the ones with plate-like morphology (compare EDS maps in Figs. 9 and 10). The layers surrounding these particles are rich in Al, Fe and Ni. Secondly, there are alternate straight lines enriched in Cr and depleted in Ni between the globular Cr carbides.

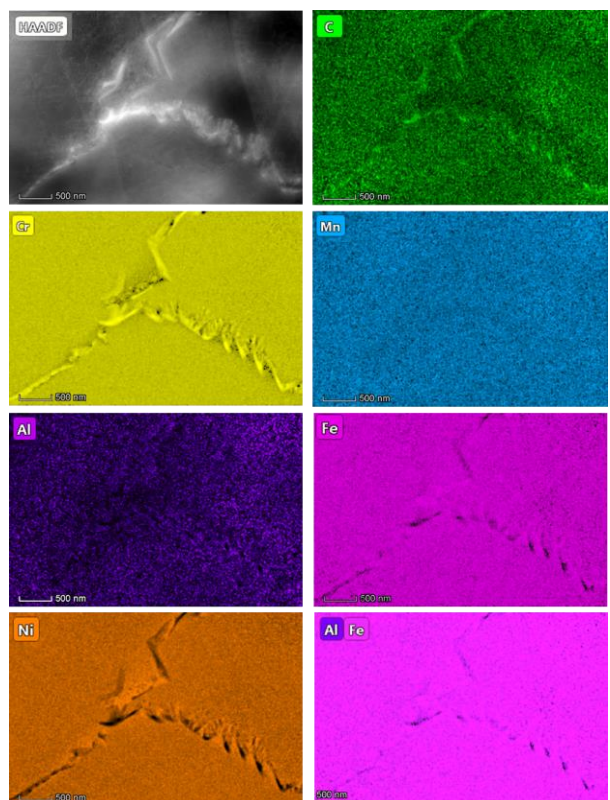

**Figure 11:** HAADF-STEM image of alloy 602 CA annealed at 403 K for 72 hours showing a faceted GB and six elemental EDS maps. Note the grain boundary plate-like particles are rich in Cr and C. Note also that the multiple layers around the particles display enrichment in Fe and Ni.

The elemental EDS maps (Cr-Ni) shown in Fig. 9b may be used to provide compositional profiles for the elements at these lines. It can be seen that these straight lines have a length of around 50 nm and appear over a width of approximately 30 nm across the GB. We note that this particular GB seems to be inclined with respect to the electron beam direction that allows quantitative observations of the spinodal-like decomposition at the GB. We are not expecting that the GB width would drastically differ from a value of about 1 nm. The HAADF intensity evolves similarly to the Ni concentration profile, however, the Cr concentration maxima coincide with Ni concentration minima suggesting these alternating lines were rich in Ni (≈ 27 at%) and Cr (≈ 12 at%), as compared to other elements. These results suggest that a chemical separation took place at the GB similar to cellular or spinodal decomposition.

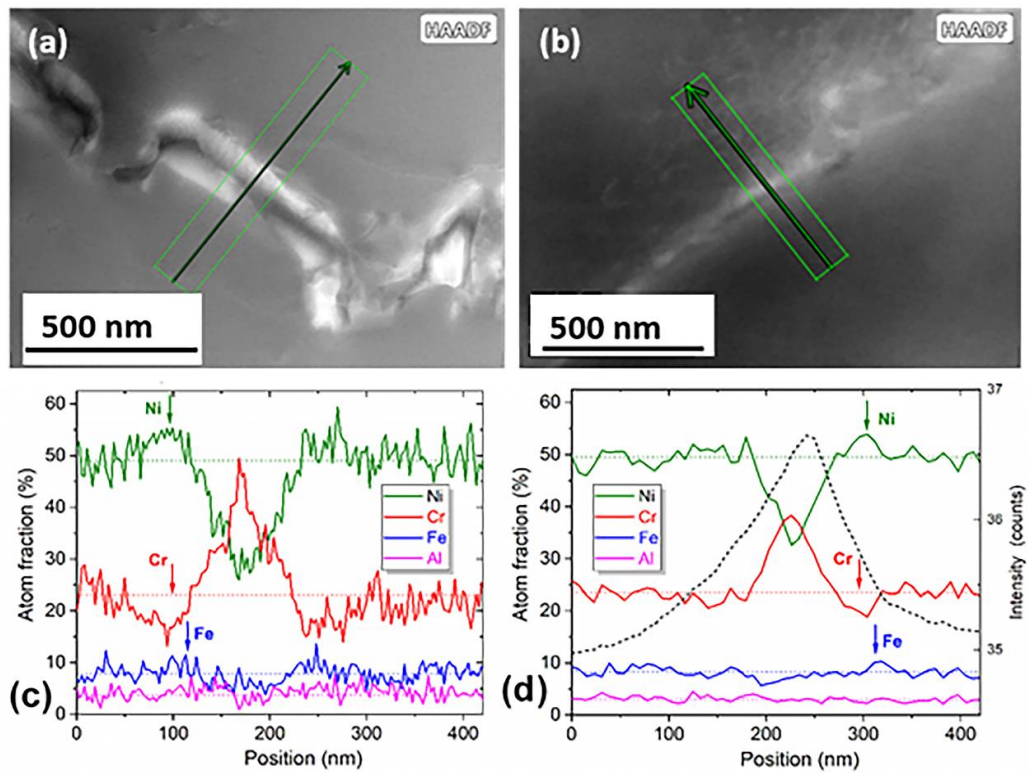

**Figure 12**: Element enrichment at the plate-like carbide precipitates after annealing at 873 K for 18 hours (a, c) and at 403 K for 72 hours (b, d). EDS map scans (c, d) over the areas indicated by green bars in (a) and (b) reveal Ni-, Fe- and sometimes Al-rich layers around the carbides. The dotted lines correspond to element concentrations reasonably far from the precipitates and the dashed line represent HAADF intensities.

Though atom probe tomography investigation and correlative microscopy would be required to quantify the chemical distribution in the segregated grain boundaries, the most intriguing aspect of the above analysis is the difference in the GB compositional variation in this alloy annealed at two temperatures in terms of GB-related precipitation, segregation and phase transformation. Based on the extensive chemical composition analysis along the grain boundaries, the complex chemical structure can be reasonably visualized, considering the schematic shown in Fig. 13 below for both, lower and higher temperature conditions. Remember, based on the literature [5] and thermodynamic calculations using the Thermo-Calc software and the TCNI8 database, the alloy 602A in solution-annealed condition (1173 K) should contain around 3% (volume fraction) of $Cr_{23}C_6$ carbides in its microstructure.

**4.2.2 GB structures at lower temperatures (<700 K)**

Below about 700 K we are observing dominantly curved and broken boundaries with plate-like carbides. Figure 11 shows a HAADF-STEM image revealing several salient features of the material annealed at 403 K for 72 hours. Besides, there are six elemental maps displayed from the region shown in the HAADF image. The GB morphology appears to be discontinuous and locally faceted. It can be seen that the GBs are covered with numerous particles which are rich in Cr and C. In contrast to the precipitates at 873 K, Fig. 10, practically no Mn enrichment is seen, Fig. 11.

The particles are surrounded by layers of several nm in thickness, enriched in Ni and to some extent in Fe. Practically no Al enrichment is seen, that is clearly different from the situation at higher temperatures, Fig. 10. The above results suggest that the particles present at the GBs possibly are $Cr_{23}C_6$ carbides surrounded by sequential segregation layers of Fe and Ni. Thus, it is implied that a strong compositional gradation exists due to discontinuous carbide precipitation and elemental segregation at the grain boundaries after annealing at 403 K for 72 hours.

The element enrichment at the precipitates as observed at lower (C-type) and higher (B-type) temperatures is compared in Fig. 12.

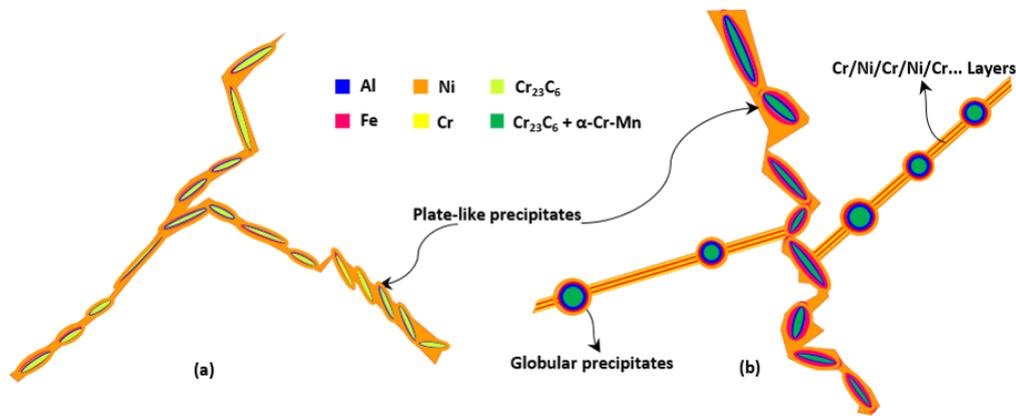

**Figure 13:** Schematic diagram showing the precipitation, elemental enrichment and the phase transformation at high-angle grain boundaries of the alloy 602 CA after annealing at two different temperatures: (a) $Cr_{23}C_6$ type carbide formation and segregation of sequential Fe- and Ni-rich layers around the carbide particles after annealing at 403 K for 72 hours and (b) grain boundary carbides transforming to $M_{23}C_6$ carbide type precipitates of different shapes on straight (globular) and hackly (plate-like) GBs. These particles coexist in equilibrium with the BCC α-Cr-Mn phase at 873 K for 18 hours and these carbides also exhibit similar enrichment layers. The straight grain boundary plane contains alternating Cr- and Ni-enriched layers resembling a cellular or spinodal decomposition.

The comparison substantiates that the precipitates are (Cr, Mn)-rich at higher temperatures and Cr-rich at lower. The Ni profiles show opposite trends to those of Cr. A Fe-rich layer at the particle/matrix phase boundary is most prominent at a higher temperature, still it is often seen at lower temperatures, too. Al layers were observed after annealing at 873 K. The Ni-rich layer is most wide and is compensated by a Cr depletion.

### 4.2.3 GB segregation, precipitation and GB phase transitions

The results of the TEM investigation shown in Figs. 9 – 12 are schematically summarized in Fig. 13 with respect to the GB structure, precipitation and element enrichment around the GB carbides as observed after annealing at lower and higher temperatures.

Following features are highlighted:

- Both straight and hackly grain boundaries are observed at higher temperatures, whereas the second type strongly dominates the GB population at lower temperatures;
- After annealing at 873 K for 18 hours, the chemical composition of the grain boundary areas is very different from that after annealing at 403 K for 72 hours. The straight boundaries are primarily decorated with globular precipitates and the hackly grain boundaries are filled with plate-like precipitates (Fig. 13b).
- The straight segments reveal alternating Ni-rich & Cr-poor / Ni-poor & Cr-rich layers as seen from the 2D projections in our TEM analysis for suitably oriented GBs, that suggests spinodal-like decomposition at 873 K (Fig. 13b). At a higher resolution one may recognize concentration patterns within the layers although the TEM technique used cannot provide tomographic 3D data;
- The element enrichment layers around the particles follow a particular sequence, first Al-rich, then Fe-rich and finally a Ni-rich layer when starting from the particle (Fig. 12). Most prominent Al- and Fe-rich layers are seen at 873 K. This behaviour is consistent with the observation of the particles in the crystalline bulk, too, and implies a separation tendency at the phase boundary.

### 4.2.4 Thermodynamic calculations

The presence of different precipitates at GBs can be interpreted on the basis of thermodynamic calculations using the Thermo-Calc software in conjunction with the Ni-alloys database TCNI8 [19]. Figure 14 shows the equilibrium volume fraction of phases present in the alloy as a function of temperature (473 – 1173 K).

At 1173 K, the temperature of solution annealing, a three-phase equilibrium of Ni-base solid solution (92.6 vol.%), $M_{23}C_6$ carbides (3.3 vol.%) and a small fraction of $Ni_{17}Y_2$ (0.5 vol.%) is predicted in agreement with the microstructure analysis reported above (Figure *2*). The plot shows that the γ′ solvus is at 1061 K and Cr-rich bcc (α-Cr) begins to form at 1009 K.

At 873 K the microstructure consists of γ (74 vol %), γ′ (15.8 vol %), α-Cr (6.3 vol %), M$_{23}$C$_6$ (3.4 vol %) and Ni$_{17}$Y$_2$ (0.5 vol %). The amounts of M$_{23}$C$_6$ and Ni$_{17}$Y$_2$ are not sensitive to the temperature variation. The calculations also predict very little solubility for Mn in M$_{23}$C$_6$ (Table 5). Since the volume fraction of M$_{23}$C$_6$ carbides is almost independent of temperature, no significant growth/dissolution of carbides during diffusion experiments can be expected in the temperature range under investigation, only some Ostwald ripening could occur.

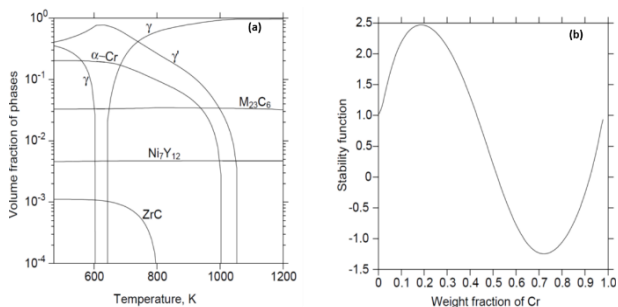

**Figure 14:** Plots showing the calculations done using the Thermo-Calc software employing the TCNI8 thermodynamic database: (a) equilibrium volume fraction of phases present in the alloy as a function of temperature (473-1173 K); and (b) the value of the stability function for α-Cr at 873 K.

The equilibrium fractions of α-Cr and γ′-type phases were found to be temperature-dependent and they increase generally with decreasing temperature. One can thus expect phase transformations to occur at the temperatures of the diffusion experiments in the initially solution-annealed alloy (at 1173 K). The completeness of the transformations and, thus, the resulting microstructure depend on the corresponding kinetics. As mentioned above, the conditions for the B- and C-regime were chosen in such a way that the bulk diffusion is limited (B-regime) or completely "frozen" (C-regime) (Table 4). Therefore, we do not expect any significant phase transformation occurring in the bulk during the diffusion experiment. However, GBs represent ways of accelerated diffusion and are known to be thermodynamically favourable for phase nucleation processes. Hence, some GB precipitation could be expected. Obviously, the rate and progress of GB phase transformations are dependent as well on temperature and time.

Thus, the precipitates visible at the GBs after annealing at 873 K for 18 hours taking into account EDS composition maps (Figure 10) are possibly a phase mixture of M$_{23}$C$_6$ + BCC α-(Cr,Mn) as sketched in Fig. 13b. Simultaneously, a phase transformation is prominent under these conditions giving rise to the appearance of alternating layers of Ni-rich/Cr-poor and Cr-rich/Ni-poor layers at the straight grain boundary segments. Although such features were observed on straight GB segments inclined with respect to the beam direction, it seems possible that curved grain boundary planes may display similar features of phase decomposition, too. However, they were not oriented properly towards the incident electron beam to be revealed clearly.

Similar first-order transitions or spinodal decompositions on a grain boundary have recently been reported for Fe-Mn binary alloys [45, 46]. Thus, the results in the present investigation clearly illustrate grain boundary segregation and the occurrence of a grain boundary phase transformation between the two annealing temperatures. The change in the precipitate shape at 873 K is probably accompanied by a prominent change of GB appearance, namely the formation of straight GBs.

Spinodal decomposition at GBs with the formation of periodic concentration fields (as observed for Ni and Cr in Fig. 9b) indicates a de-mixing tendency between Ni and Cr atoms at 873 K for the given GB composition. It is worth remembering that no Ni nor Cr concentration bands were observed in the bulk for the same conditions. On the one hand, the absence of spinodal decomposition in the bulk can be due to limited bulk diffusion at 873 K as discussed earlier. On the other hand, the thermodynamic activities of the elements constituting the studied alloy are different for the given conditions as compared for the bulk and for the GB to equilibrate the difference due to GB energy and to have equal chemical potentials [47]. Therefore, the concentration dependence of the thermodynamic activities of the elements in the bulk and in the GB should be different. Hence, the GBs can theoretically undergo phase transformations that are not expected (observed) in the bulk. In such a case spinodal decomposition in GBs can occur without previous/simultaneous/subsequent spinodal decomposition in the bulk.

Here we performed simple estimates for the possibility of spinodal-like decomposition. The value of the stability function for bcc α-Cr at 873 K is positive (QF$_{α-Cr}$ = 0.903), indicating that it is outside the spinodal regime (Fig. 14b). However, in the neighborhood of $M_{23}C_6$ this phase is depleted in chromium that shifts it to the spinodal regime, where the value of the stability function becomes negative. This could explain the presence of the two layers of bcc phase of differing chemistry, but depleted of chromium, along the grain boundary where $M_{23}C_6$ is seen. Then the following scenario may be suggested. A high-angle grain boundary in an FCC alloy is a 2D defect and the corresponding coordination number is smaller, 8 to 10. We may qualitatively treat this situation as a 3D bcc phase (with the coordination number 8). The carbide formation at GBs induces a Cr depletion and Cr redistribution along the GB. Then in Cr-poor areas, a spinodal-like instability may occur and alternative Ni-rich/Cr-rich layers are formed. This explains the appearance of such layers between the carbides.

Table 5: Calculated equilibrium constitution of the alloy at 873 K (in at.%). The Ni amount is balanced.

| Phase | Cr | Fe | Al | C | Ti | Mn | Si | Cu | Zr | Y |
|---|---|---|---|---|---|---|---|---|---|---|
| $\gamma$ | 21.23 | 11.55 | 1.35 | 1.9E-4 | 7.83E-3 | 8.20E-2 | 9.22E-2 | 9.92E-3 | 1.93E-2 | 1.59E-10 |
| $\gamma'$ | 6.87 | 4.14 | 7.82 | 1.86E-4 | 0.799 | 4.45E-2 | 3.71E-3 | 1.63E-2 | 0.422 | 1.67E-10 |
| $\alpha$-Cr | 97.55 | 2.12 | 1.92E-4 | 1.94E-7 | 2.16E-8 | 2.31E-2 | 3.39E-6 | 1.22E-10 | 1.69E-8 | 2.05E-10 |
| $M_{23}C_6$ | 92.02 | 1.43 | 0 | 5.67 | 0 | 8.26E-7 | 0 | 0 | 0 | 0 |
| $Ni_{17}Y_2$ | 0 | 0.541 | 3.81 | 0 | 0 | 0 | 0 | 0 | 0 | 15.8 |

To be able to give a quantitative theoretical description of the spinodal decomposition, the Gibbs energy of the GB solid solution should be a smooth function of the solute content having at least three extrema. Classically, the regular solution model can be used which, for a certain value of interaction energy, results in demixing of the solution constituents. Considering the bulk as ideal solution and the GB as regular solution, the Fowler-Guggenheim GB segregation isotherm can be obtained [48]. This isotherm was recently used to describe spinodal decomposition in GBs of binary Fe-Mn binary alloys [45]. Therefore, the thermodynamic description of GB spinodal decomposition in our Ni-Cr-Fe-Al alloy is a quite complex problem and will be published elsewhere.

#### 4.2.4 Correlation between GB diffusion and GB structures

As we mentioned, we cannot state any *direct* relationship between the GB structure as seen in high-resolution TEM and the corresponding GB diffusion rate. The tracer diffusion technique by mechanical sectioning is an inherently integral method in which the tracer distributions created by individual interfaces are averaged. Although the averaged grain size in the material is about 45 μm, a broad distribution of the grain sizes is seen with some grains of more than 200 μm in diameter (Fig. 1). Since the penetration depths are less than 100 μm (Figs. 5a and 5b) the deepest branch of concentration profiles correspond to diffusion along (probably not a large number of) individual "fastest" GBs, while the first steepest branch represents the contribution of the "slowest" interfaces. Note that these slowest GBs are still short-circuits for diffusion, see the discussion above. The transition between the two branches is relatively sharp supporting a view of two families of GBs with distinct diffusivities.

### 4.3. Impact of GB structure transitions on GB diffusion

The present radiotracer study supports strongly the interpretation that the two distinct families of short-circuit paths found in the material are represented by two types of high-angle grain boundaries, namely by straight and curved boundaries, that are associated with the two distinct types of structures and structural transformations. This view is advocated both by the relative fractions of occurrence of the two GB types and the specific features of the temperature-induced transformations. The dense segregation of carbides at HAGBs with higher excess energy density was also reported by Bennett et al. [49] and Trillo et al. [50,51] after annealing of stainless steel at 943 K. The arrangement (II), for details see Section 4.1, with independent diffusion paths along two distinct families of high-angle GBs seems hence to provide the most reasonable explanation for the present experimental observations and agrees with the results of a careful microstructure examination.

The determined GB diffusion parameters, i.e. the triple products $P_F$, $P_S$ (left ordinate, B-type kinetics) and $D_{gb}$ (right ordinate, C-type kinetics), for Ni, Fe and Cr grain boundary diffusion are plotted in Fig. 8a as functions of the inverse temperature and are plotted against the inverse homologous temperature in Fig. 8b.

Let us first analyse the diffusion data for Cr and Fe. Figure 8 reveals that the grain boundary diffusivities of fast paths, i.e. the triple products $P_{F-Cr}$ for $^{51}$Cr GB diffusion at 873 and 823 K (closed blue stars) are higher than the findings of Ćermăk [11] for nearly the same composition, a Ni-19Cr-10Fe alloy (blue solid line). The triple product $P_{S-Cr}$ (open blue stars) of the slow diffusion path is, however, closer to the findings of Ćermăk [11]. In the case of $^{59}$Fe diffusion however, the $P_{F-Fe}$ (closed inverted green triangles), seems to be closer to the estimated triple product for the Ni-19Cr-10Fe alloy (green solid line). The triple products $P_{S-Fe}$ for the slow path (open inverted green triangles) are one order of magnitude lower than for the Ni-19Cr-10Fe alloy. Since GB diffusion, in this case, was investigated at two temperatures only, the corresponding Arrhenius fits would not be reliable and they are not reported here.

Ćermăk [11] did not report the existence of two types of short-circuit paths for all alloys investigated. The possible reason being that the alloys investigated had only Ni, Cr and Fe as major constituting elements that allowed only single-phase formation without precipitation or segregation. However, a careful inspection of the published penetration profiles in Ref. [11] supports the present observations, since further branches are clearly seen between the potential volume diffusion and fast grain boundary diffusion contributions. These are the features which were obviously seen for the penetration profiles measured in the present work.

In Fig. 8a and b, the corresponding values of the Ni grain boundary diffusion triple products are plotted against the inverse temperature and almost linear Arrhenius dependencies are seen. The triple products for the fast, $P_{F-Ni}$ and slow, $P_{S-Ni}$ branches are found to have the following temperature dependencies,

$$P_{F-Ni} = 3.2^{+34.9}_{-2.9} \times 10^{-15} \exp\left(-\frac{128 \pm 17 \, kJ \, mol^{-1}}{RT}\right) m^3 \, s^{-1} \quad (10)$$

$$P_{S-Ni} = 7.4^{+658}_{-0.08} \times 10^{-7} \exp\left(-\frac{285 \pm 31 \, kJ \, mol^{-1}}{RT}\right) m^3 \, s^{-1} \quad (11)$$

Formally, the activation enthalpy of $^{63}$Ni diffusion along the fast short-circuit paths (F) in the B-type regime, equation (10), turned out to be almost equal to that of nickel diffusion in high-purity nickel (material A in Ref. [15]). The presence of impurities (solutes) reportedly diminishes the rate of GB diffusion of both, solvent [52] and solutes [15,23,53]. The absolute values of the triple product $P_F$ for the 602CA alloy are lower than those for high-purity Ni and higher than for non-pure Ni (material B in Ref. [15], Fig 8a). In view of the abundance of carbides at grain boundaries in the 602CA alloy, Figs. 2, 3 and 9–11 one would expect a retardation of the GB diffusion rate with respect to pure Ni.

The contribution of the slow short-circuit paths was seen up to depths of about 20 μm (Fig. 5), while that of the fast paths was detected up to depths of 100 μm and more. Since the average grain size is about 46 μm, all "slow" interfaces crossing the external surface contribute to the corresponding diffusion flux, while atomic transport over several connected "fast" boundaries should correspond to the deeper branch. Thus, parallel and independent contributions of the two different types of grain boundaries provide a reasonable origin of the observed shape of the penetration profiles.

As it was mentioned, the radiotracer diffusion technique is an integral method while TEM investigations are highly local. Thus, we cannot unambiguously prescribe the diffusion properties, i.e. slow or fast, to the given types of high-angle GBs found by TEM inspection. However, an analysis advocates strongly that the curved hackly GBs, Figs. 4, 10, 11, represent the fast short-circuit paths in view of highly distorted structures with probably an increased associated excess free volume. The straight GBs with globular carbides and detectable segregation of Cr (or associated decomposition in alternating Cr- and Ni-rich layers along the GB plane between the carbides), Fig. 9, represent most probably the slow paths for the diffusion of Ni atoms. Such GBs dominate grain boundary diffusion transport at lower temperatures in the C-type kinetic regime, too. Therefore, those diffusion results need a special consideration.

### 4.4. Kinetic evidences for metastable grain boundary transitions

The present measurements of Ni diffusion in the 602CA alloy under C-type kinetics conditions yielded unexpected results, Fig. 8a. The GB diffusion coefficient of 4.89×10$^{-17}$ m$^2$/s at 673 K is similar to the results for Ni self-diffusion in low-purity Ni (material B in Ref. [15]) as seen in Fig. 8a, cf. magenta open circles and the corresponding green triangles up. However, the diffusion coefficients $D_{gb}$ determined at temperatures below 673 K are unusually high and their values do not practically depend on the temperature. The tentatively determined activation enthalpy would be as small as 3.5 kJ/mol, a value which can hardly be attributed to any thermally activated process.

By extrapolating the Arrhenius dependence found for the triple product of fast (F) diffusion paths, Eq. (10), to a lower temperature of 673K, the product of the effective GB width and the segregation factor $s$ can be evaluated as

$$s\delta = \frac{P}{D_{gb}} \qquad (12)$$

The product $s\delta$ was found to be 7.2±0.3 nm at 673 K. If one assumes that $s \approx 1$ (see the discussion above, a slight segregation of Cr in Fig. 9 corresponds to the segregation factor less than 2), the determined value of $\delta$ would be drastically different from the widely accepted GB width of about 0.5 nm, as reported for various binary alloy systems and pure metals [23]. As a hypothetical explanation one may propose that the Ni segregation factor is not about 1, but equals to about 10. However, since the bulk concentration of Ni is 62 wt. %, see Table 1, the segregation factor is definitely less than 2, which would correspond to a pure Ni interface.

An alternative explanation is required and we propose that it is a *transition to a metastable GB structure* which affects the GB diffusion rates and hinders the applicability of Eq. (12). Parts of the scientific community, see e.g. [54, 55] refer to GB structure transitions as so-called "complexions". In those terms, one would use a term of *metastable complexion* for the transition observed in the present work. The Dillon-Hammer [54,55] classification quite elegantly describes the occurrence of grain boundary complexions in dependence of structural disorder, adsorption, or boundary thickness at equilibrium in alumina-based systems.

However, we would prefer to use the term GB transition or transformation between thermodynamically stable (or metastable or conditionally stable – if the existence of an interface is stabilized by external conditions) states.

The present results can be explained by different grain boundary states in the given system due to localized alterations of the alloy composition (basically the segregation level and the precipitations) and the microscopic degrees of freedom of the internal interfaces. Fast diffusion paths are thought to be represented by high-angle GBs with enriched curved discontinuous plate-like carbides (Figs. 4 and 10) that create large localized structural and compositional disorder (Figs. 10 – 12) similar to a combination of complexions III & V [54,55]. The slow diffusion paths correspond to the boundaries with tiny globular carbides connected by alternating Ni-Cr enriched straight lines, Fig. 9. Especially at low temperatures corresponding to the C-type regime, the discontinuous faceted GBs provide likely fast paths for diffusion.

The combined results of the B- and C-type regime measurements, when referenced to GBs with the same – i.e. highly curved and broken – appearance, are evidently incompatible and we propose that an incomplete and continuous relaxation of elastic strains induced by carbides at low temperatures (under the C-type GB diffusion kinetics) was responsible for this behaviour. One may define the corresponding GB state as a *continuous GB transition* or *metastable GB state*. In such a case, the formally determined diffusion coefficients in the C-type regime do not represent material's constants and are purely effective values which depend on the given conditions, including the diffusion time at the given diffusion temperature.

This assumption was checked by performing time-dependent diffusion measurements at 403 K for short (16 hours) and long (87 hours) diffusion times. The measured penetration profiles are shown in Fig. 15. A strong variation of the effective diffusion coefficients (which are proportional to the slopes of the linear fits) with the annealing duration is obvious. Hence, we conclude that a continuous relaxation (continuous GB phase transition) might occur in the range of lower temperatures, between 403 K and 673 K. Furthermore, some fraction of carbides could be formed during diffusion annealing affecting the measured values and making them time-dependent. In Fig. 15 the measured penetration profiles are rescaled for the conditions of C-type GB diffusion since a formal Gaussian type solution is expected.

The time scales of GB relaxation (including the relaxation of transformation-induced strains) and of GB diffusion measurements are basically different since the relaxation of local strains requires atomic jumps in the neighbouring crystalline volume while jumps in the distorted GB-related area contribute to the short-circuit GB diffusion. The diffusion kinetics in the grain interiors is retarded with respect to GB diffusion. In fact, the existence of metastable GB states with relaxation times, at lower temperatures, significantly exceed those for GB diffusion; such a conclusion was already reported for enhanced GB diffusion in severely plastically deformed materials [15,38] in terms of a 'lock-in' model [56].

In the context of thermal equilibrium in GBs, Han et al. [57] discussed the non-Arrhenius temperature dependence of intrinsic GB mobility. Several atomistic model simulations support some of the experimental observations that the intrinsic GB mobility increases with decreasing temperature. The present results suggest that this theory of non-thermal (or non- Arrhenius) structural phase transitions can be extended to GB relaxation in the alloy under investigation which reveals the existence of GB complexions.

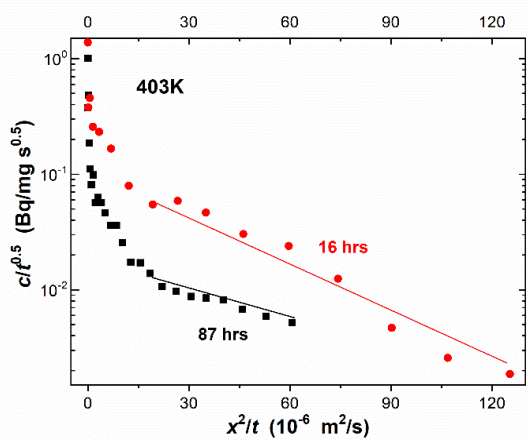

**Figure 15**: Ni tracer diffusion penetration profiles for an annealing temperature of 403K, corresponding to the C-type regime for two exposure times: 87 and 16 hours. The corresponding slopes of profiles (indicated by the straight lines) are significantly different.

The localized elastic strains at GBs and in the neighbouring regions in alloy 602A annealed at two different temperatures were mapped using nano-beam diffraction methods. The strain maps and the detailed precipitation and segregation data were correlated and the corresponding detailed investigation will be reported elsewhere [58]. The results reveal the existence of high residual elastic strains localized at GB which presumably accelerate GB diffusion, as already reported for a severely deformed alloy [40]. As temperature increases, the elastic strains relax continuously, thus affecting the diffusion measurements at lower temperatures under C-type conditions. As a result, almost temperature-independent effective diffusion coefficients are measured in the C-type kinetic regime, Fig. 5a and Fig. 8. Note that with increasing temperature the relaxation of the transformation-induced strains is more completed within the time window for the given diffusion measurements and relatively lower values of $D_{gb}$ are determined at higher temperatures, resulting in almost temperature-independent diffusion coefficients in the Arrhenius diagram, Fig. 8.

At even higher temperatures in the B-type kinetics regime, the strain relaxation time becomes significantly shorter compared to the diffusion times, grain boundary diffusion experiment proceeds at almost equilibrium conditions and diffusion-time independent (relaxed) diffusion parameters are measured.

## 5. Conclusions

Tracer grain boundary diffusion in a coarse-grained Ni-Cr-Fe 602CA alloy was measured in formal Harrison's B- and C-type kinetic regimes applying $^{63}$Ni, $^{59}$Fe and $^{51}$Cr radioisotopes. The iron and chromium diffusion data measured under the B-type kinetic conditions are in good agreement with the literature values reported for alloys with similar compositions.

The B-regime tracer diffusion measurements reveal unambiguously the existence of two branches, i.e. slow and fast short-circuit diffusion paths, which can be convincingly explained in terms of two distinct families of high-angle grain boundaries with significantly different precipitation, phase transformation, segregation morphologies and interface appearance. HAADF-STEM investigations of the alloy annealed at 873 K for 18 hours, i.e. under the conditions corresponding to the B-type kinetic regime of GB diffusion measurements, reveal the following:

1. The fast diffusion is likely to correspond to the structurally disordered GBs. These GBs are curved and hackly, covered by plate-like precipitates containing $(Cr,Mn)_{23}C_6$ type carbides coexisting with a Cr-Mn-rich phase in addition to sequential segregation layers of Al, Fe and Ni around them.
2. The slow diffusion paths are due to straight grain boundaries composed of globular carbides additionally showing alternating layers enriched in Cr and Ni, similar to the resulting microstructure expected after spinodal-like decomposition.

The Ni diffusion rates at relatively low temperatures in the formal C-type kinetics regime showed an anomalous character being almost temperature independent. This was interpreted as a continuous strain relaxation and structural transition of the grain boundary, i.e. as a metastable GB state. HAADF-STEM investigations of the alloy grain boundary, annealed at 403 K for 72 hours, i.e. under the conditions corresponding to the C-type kinetic regime of GB diffusion measurements, revealed plate-like $Cr_{23}C_6$ type carbides with segregation of Al-, Fe- and Ni-rich layers around them, probably as a depletion zone. The GB structure and the GB kinetic properties also reveal intricate temperature and time dependencies.

The correlative atomistic structure − kinetic properties measurements provide unique insights into the interplay of GB structure transitions and the resulting diffusion behaviour in the Ni-base alloy with respect to GB-related precipitation, phase transformation and segregation.

**Acknowledgments**

Sai Rajeshwari would like to thank Deutscher Akademischer Austauschdienst (DAAD) for providing a scholarship to perform tracer diffusion studies at WWU Muenster. S. Sankaran thanks Alexander von Humboldt Foundation for his research stay in Germany through an AvH fellowship for experienced researchers. Partial financial support by Deutsche Forschungsgemeinschaft (DFG) via a research project DI 1419/16-1 is acknowledged. The DFG is further acknowledged for funding our TEM equipment via the Major Research Instrumentation Programme under INST 211/719-1 FUGG.